\documentclass[sigconf]{acmart}
\AtBeginDocument{%
  \providecommand\BibTeX{{%
    \normalfont B\kern-0.5em{\scshape i\kern-0.25em b}\kern-0.8em\TeX}}}

\copyrightyear{2022}
\acmYear{2022}
\setcopyright{acmcopyright}
\acmDOI{XXXXXXX.XXXXXXX}

\acmConference[SIGIR '22]{Proceedings of the 45th International ACM SIGIR Conference on Research and Development in Information Retrieval}{July 11--15, 2022}{Madrid, Spain.}

\acmBooktitle{Proceedings of the 45th International ACM SIGIR Conference on Research and Development in Information Retrieval (SIGIR '22),
July 11--15, 2022, Madrid, Spain} 
\acmPrice{15.00}
\acmISBN{978-1-4503-8732-3/22/07}
\acmDOI{10.1145/3477495.3531801}

\usepackage{graphicx}
\usepackage{subcaption}
\usepackage{multirow}
\usepackage{hyperref}

\begin{document}
\fancyhead{}

\title{Empowering Next POI Recommendation with \\Multi-Relational Modeling}

\author{Zheng Huang}
\affiliation{%
  \institution{University of Virginia}
  \city{Charlottesville}
  \state{Virginia}
  \country{USA}
  \postcode{22904}}
\email{zh4zn@virginia.edu}

\author{Jing Ma}
\affiliation{%
  \institution{University of Virginia}
  \city{Charlottesville}
  \state{Virginia}
  \country{USA}
  \postcode{22904}}
\email{jm3mr@virginia.edu}

\author{Yushun Dong}
\affiliation{%
  \institution{University of Virginia}
  \city{Charlottesville}
  \state{Virginia}
  \country{USA}
  \postcode{22904}}
\email{yd6eb@virginia.edu}

\author{Natasha	Zhang Foutz}
\affiliation{%
  \institution{University of Virginia}
  \city{Charlottesville}
  \state{Virginia}
  \country{USA}
  \postcode{22904}}
\email{nfoutz@virginia.edu}

\author{Jundong	Li}
\affiliation{%
  \institution{University of Virginia}
  \city{Charlottesville}
  \state{Virginia}
  \country{USA}
  \postcode{22904}}
\email{jl6qk@virginia.edu	}




\begin{abstract}
  With the wide adoption of mobile devices and web applications, location-based social networks (LBSNs) offer large-scale individual-level location-related activities and experiences. Next point-of-interest (POI) recommendation is one of the most important tasks in LBSNs, aiming to make personalized recommendations of next suitable locations to users by discovering preferences from users' historical activities. 
  Noticeably, LBSNs have offered unparalleled access to abundant heterogeneous relational information about users and POIs (including user-user social relations, such as families or colleagues; and user-POI visiting relations). Such relational information holds great potential to facilitate the next POI recommendation. 
  However, most existing methods either focus on merely the user-POI visits, or handle different relations based on over-simplified assumptions while neglecting relational heterogeneities. 
  To fill these critical voids,
  we propose a novel framework, MEMO, which effectively utilizes the heterogeneous relations with a multi-network representation learning module, and explicitly incorporates the inter-temporal user-POI mutual influence with the coupled recurrent neural networks. Extensive experiments on real-world LBSN data validate the superiority of our framework over the state-of-the-art next POI recommendation methods.

\end{abstract}





\maketitle

\section{Introduction}
The era of information explosion has witnessed rapid developments and adoptions of mobile devices and web applications. As a result, location-based social networks (LBSNs) have recorded rich activities at different \textit{points-of-interests (POIs)}, such as shopping malls, restaurants, or gyms. Next POI recommendation represents one of the most important tasks in LBSNs \cite{li2018next,zhao2020go}. It aims to suggest the next suitable location(s) for each person based on his/her preference that can be learned from his/her prior information on LBSNs, such as user profile, social relationship, and historical activities.

Prior research has performed next POI recommendation by capturing users' preferences from their trajectories, i.e., users' temporally ordered location information at different POIs. For instance, sequential statistical models (e.g. Markov chain) \cite{baumann2013influence,gambs2012next} are widely used. In addition, graph-based methods \cite{xie2016learning, he2020lightgcn} learn both user and POI representations for recommendation by modeling each user-POI visit as an edge in a graph. Another line of research \cite{feng2018deepmove,zhao2020go,zhu2017next} applies recurrent neural networks (RNNs) based models to capture users' preferences by modeling users' trajectories as sequences. More recent studies adopt self-attention \cite{vaswani2017attention} to capture  long-term dependencies and spatio-temporal correlations among POIs \cite{lian2020geography,luo2021stan}.

\begin{figure}[!t]
  \vspace{-2mm}
  \centering
  \includegraphics[width=1.0\linewidth, height=0.91in]{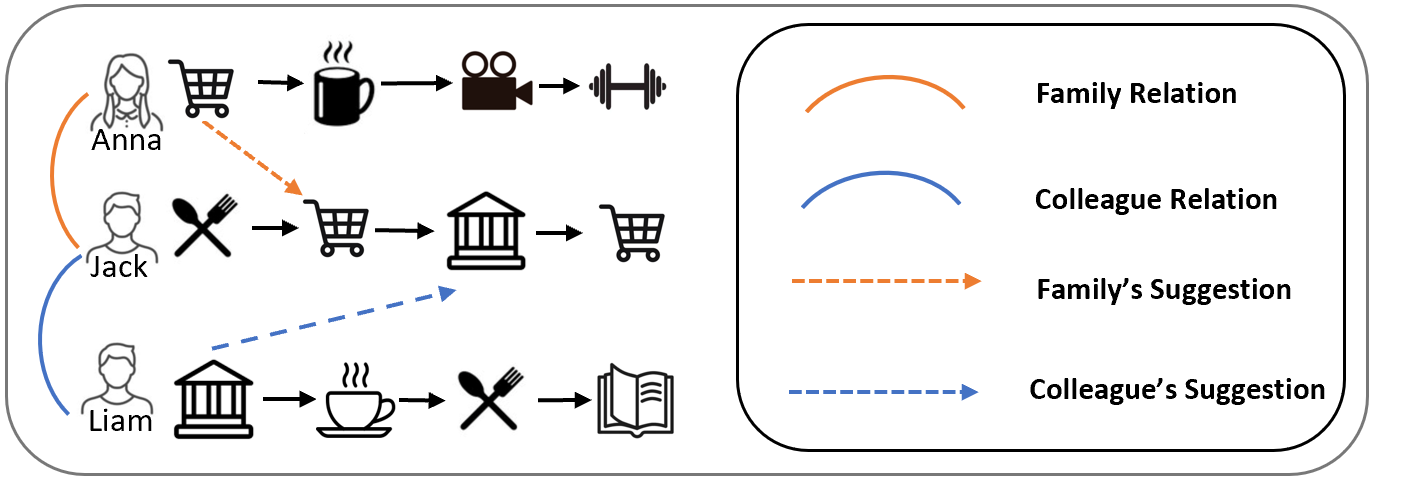}
    \vspace{-4mm}
  \caption{Heterogeneous social relationship: Jack takes sister Anna's recommendations on shopping malls and colleague Liam's recommendations on training institutions.
  }
  \vspace{-5mm}
  \label{fig:location}
\end{figure}

Despite the success of these methods, next POI recommendation on LBSNs remains a daunting task due to the following challenges: 1) users' preferences are complex and often bear strong connections with different types of social relations~\cite{mcpherson2001birds,tang2012mtrust}. As indicated in \cite{massa2007trust}, different social relations affect users differently. For example, users seek recommendations on shopping malls from family members, whereas recommendations on training institutions from colleagues (Fig.~\ref{fig:location}). These heterogeneous social relations, together with the user-POI relations formed by the users' visits to POIs, bring great challenges to effectively utilize the rich information embedded. Most existing methods \cite{sun2020go,yang2017bridging, lim2020stp, xie2016learning} either focus solely on the user-POI relations, or assume the node representations in different relation-specific networks are within the same latent space. 2) Capturing the crucial, inter-temporal, mutual influence between the users and POIs remains challenging. In the user-POI relations, users' preferences and POIs' latent status (e.g., reputation) reciprocally influence each other over time. As shown in Fig.~\ref{fig:location}, if Anna left a positive comment about a shopping mall that she recently visited, or recommended it to others, then the reputation and popularity of this POI may be elevated. In turn, the experience and reputation with this POI will affect Anna's future visitation.

To address these challenges, we propose a novel framework, \underline{M}ulti-R\underline{e}lational \underline{Mo}deling (MEMO) for next POI recommendation. First, to utilize the diverse user-user social relations (e.g., families or colleagues) and user-POI relations, we develop a multi-relational modeling module based on multiple graph convolutional networks (GCNs) \cite{hamilton2017inductive} and self-attention \cite{kang2018self}. Specifically, we map each type of relation to a corresponding network to learn the relation-specific representations; and then adopt a self-attention mechanism to aggregate the user representations from the different relation types.
Next, to capture the mutual influence between the users and POIs over time, we design a user-POI mutual influence modeling component based on coupled recurrent neural networks (RNNs) that mutually update each other's representations \cite{kumar2019predicting}.
  
In summary, this research contributes to the literature along three dimensions. One, it extends the extant studies on next POI recommendation by accounting for the critical, yet overlooked, heterogeneous relations, particularly heterogeneous social relations among users, and inter-temporal user-POI mutual influence. Two, this research proposes a novel framework, MEMO, to incorporate the heterogeneous relations with a multi-network representation learning module, and to capture the inter-temporal user-POI mutual influence with the coupled RNNs. Three, it substantiates the proposed framework via the experiments on large-scale real-world data and comparisons with the state-of-the-art baselines.


\begin{figure*}[h]
  \centering
  \vspace{-3mm}
  \includegraphics[width=0.8\linewidth, height=2.6in]{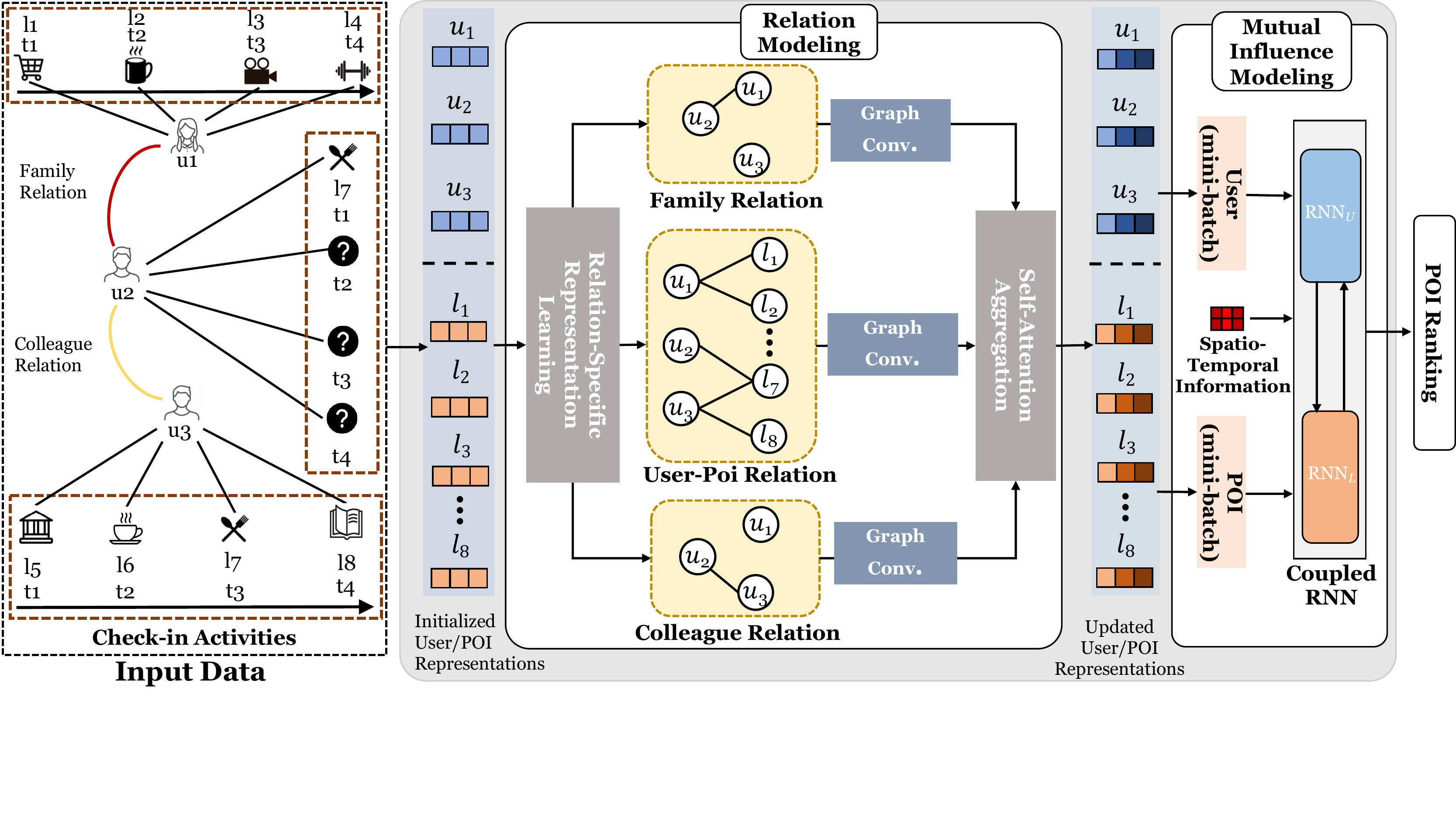}
  \vspace{-14mm}
  \caption{Overview of \emph{MEMO} for next POI recommendation.}
  \vspace{-4mm}
  \label{fig:dmgcn_framework}
\end{figure*}

\section{Problem Statement}
Let $\mathcal{U} = \{u_1, u_2,...,u_U \}$  be a set of \emph{U} users, and $\mathcal{L} = \{l_1, l_2, ..., l_L \}$ be a set of \emph{L} POIs. Each POI $l \in \mathcal{L}$ is associated with a geo-coordinate $(\text{longitude}, \text{latitude})$ indicating its geographical location. 

\noindent\textbf{Trajectory.}
The trajectory of each user $u_i \in \mathcal{U} $ is a time-ordered sequence: $C(u_i) = \{c_1(u_i), c_2(u_i),..., c_K(u_i)\}$. Each element $c_k(u_i)$ 
can be represented with a tuple $(l_k, t_k)$, where $l_k$ is the geo-coordinate of the POI visited by user $u$ at timestamp $t_k$. 

\noindent\textbf{User-User Social Relations.}
Let $\mathcal{R} = \{r_1, r_2, ..., r_P\}$ denote a set of ${P}$ social relations among the users. We define a user-user social relation as a triple $(u_i, u_j, r_p)$, where $u_i, u_j \in \mathcal{U}$, $r_p \in \mathcal{R}$, indicating that user $u_i$ and $u_j$ have a social relation of type $r_p$, which is assumed to be symmetric for simplicity. Note that there may exist multiple types of social relations between two users.

\noindent\textbf{Next POI Recommendation.}
At each timestamp $t$, the next POI recommendation task takes each user's trajectory from timestamps $1$ to $t$ and different types of social relations as the input, and recommends a list of the suitable POIs with the highest prediction scores for each user to visit at next timestamp $t+1$. 

\section{The Proposed Framework}
We now introduce the proposed framework, MEMO, consisting of two main components (Fig. \ref{fig:dmgcn_framework}): 
1) a \textbf{relation modeling} module to effectively utilize the heterogeneous relations, and 
2) a \textbf{user-POI mutual influence modeling} module to capture the mutual influence between the users and POIs over time.


\subsection{Relation Modeling}
To capture each user's preferences for the next POIs, MEMO leverages the heterogeneous relations, including multiple user-user social relations and user-POI relations collected from the LBSNs. Specifically, to model $P$ different types of user-user social relations among the user set $\mathcal{U}$, we enlist $P$ networks $\mathcal{G}_1,...,\mathcal{G}_P$, each with its corresponding symmetric adjacency matrix $\mathbf{A}_1,...,\mathbf{A}_P$, where $\mathbf{A}_p \in \mathbb{R}^{U\times U}$ and $\mathbf{A}_p[i,j]=\mathbf{A}_p[j,i]=1$ if there is a type $p$ social relation between user $u_i$ and $u_j$. Similarly, the user-POI network $\mathcal{G}_C$ corresponds to an adjacency matrix $\mathbf{A}_C\! \in\! \mathbb{R}^{(U+L)\times(U+L)}$ that captures whether a user has visited a specific POI or not. 
In total, there are $P +1$ types of relations, including $P$ types of user-user social relations and 1 type of user-POI relations.

\noindent\textbf{Relation-Specific Representation Learning.}
To accommodate the diverse types of relations, we leverage a relation-specific representation learning module to map the nodes into a latent representation space corresponding to each relation separately. First, for each node $v_i$, a general embedding $\mathbf{x}_i$ 
is initialized randomly. 
We then use a relation-specific transition function $\Phi_p(\cdot)$ to map each $\mathbf{x}_i$ to a new embedding $\mathbf{x}_i^p$ corresponding to the $p$-th relation type for any $p\in [P+1]$:  
$\mathbf{x}_i^p = \Phi_p(\mathbf{x}_i)$.
Based on $\mathbf{x}_i^p$, we learn a relation-specific representation $\mathbf{h}_i^p$ for each node $v_i$ by aggregating the embeddings of the neighbors on each network $\mathcal{G}_p$, which is implemented by a GCN \cite{hamilton2017inductive} layer with an attention mechanism \cite{velivckovic2017graph,yang2016hierarchical}.

\noindent\textbf{Aggregation over Different Relation Types.}
As discussed in \cite{hu2020heterogeneous}, one should not simply assume homogeneous relations or nodes representations across the relation-specific networks within the same latent space. 
Inspired by \cite{hu2020heterogeneous}, we hence utilize a self-attention mechanism \cite{vaswani2017attention} to aggregate all relation-specific representations of each node into a common latent space. This allows us to effectively capture each user's preference embedded in the different types of relations. Specifically, given each node $v_i$, and all its representations in the different networks, we map each representation corresponding to the $p$-th ($p\in [P+1]$) relation type $\mathbf{h}_i^p$ into a Key vector $\mathbf{k}_i^{p}=\mathbf{W}_p^K\mathbf{h}_i^p$, a Query vector $\mathbf{q}_i^{p}=\mathbf{W}_p^Q\mathbf{h}_i^p$, and a Message vector $\mathbf{m}_i^{p}=\mathbf{W}_p^M\mathbf{h}_i^p$, respectively. Here $\mathbf{W}_p^Q,\mathbf{W}_p^K,\mathbf{W}_p^M$ are the trainable parameters used for the above linear projections. 
We then calculate the attention weight between each pair of relation types $(p_1,p_2)$ with the similarity between the $p_1$-th Query vector and the $p_2$-th Key vector:
$
\alpha(p_1, p_2)= \text{Softmax}_{\forall p_2\in[P+1]}({\mathbf{k}_i^{p_2}\mathbf{q}_i^{p_1} }/{\sqrt{d}}),
$
where $d$ is the representation dimension. Note that different from the ordinary attention mechanisms \cite{velivckovic2017graph} that assume the same latent space for the input representations, the above linear projections in our framework are relation-specific, hence allowing us to aggregate the relation-specific representations from different latent spaces. We then aggregate the representations from the different relation types as
$
{\mathbf{h}}_i = \text{MLP}([\Tilde{\mathbf{h}}_i^{1}|| \Tilde{\mathbf{h}}_i^{2} || ...||\Tilde{\mathbf{h}}_i^{P+1}]), 
$ where $||$ is a concatenation operation, $\text{MLP}(\cdot)$ is a multilayer perceptron, and $\Tilde{\mathbf{h}}_i^{p}=\mathop{\oplus}\nolimits_{p'\in [P+1]} \{\alpha(p, p')\cdot \mathbf{m}_i^{p'}\}$ is the representation that aggregates the messages from all relation types to the $p$-th relation type, with $\oplus$ denoting the element-wise sum.

\subsection{User-POI Mutual Influence Modeling}

In the user-POI relations, the users' and POIs' latent status may be mutually influenced by each other over time. We hence need to update the users' and POIs' representations to capture such a mutual influence. Specifically, we deploy the coupled RNNs \cite{kumar2019predicting} consisting of 
a user RNN ($\text{RNN}_U$) and a location RNN ($\text{RNN}_L$) to respectively learn the users' and POIs' representations at each timestamp. 

Here, $\text{RNN}_U$ incorporates the POIs' representations to update the users' representations, and vice versa. Suppose that user $\emph{u}$ visited POI $\emph{l}$ 
at timestamp ${t}$ in $C(u)$, user $u$'s representation at timestamp ${t+1}$ is influenced by location $l$'s representation at timestamp $t$, formally derived as:
$
  \mathbf{ h}_u^{t+1} = \sigma (\mathbf{W}_1^U\mathbf{h}_u^t + \mathbf{W}_2^U\mathbf{h}_l^t+ \mathbf{W}_3^U\mathbf{z}^{\triangle t}+ \mathbf{W}_4^U\mathbf{z}^{\triangle d}).
$
The matrices $\mathbf{W}_1^U,...,\mathbf{W}_4^U$ are the parameters of $\text{ RNN}_U$. For simplicity, we use $\mathbf{h}_u^{t+1}$ and  $\mathbf{h}_u^t$ to respectively denote user $u$'s representations at timestamp $t+1$ and $t$; and $\mathbf{h}_l^t$ the representation of location $l$ at timestamp $t$. Here, $\mathbf{z}^{\triangle d}$ and $\mathbf{z}^{\triangle t}$ are the representations incorporating the spatio-temporal information to facilitate next POI recommendation (details later in this section). 

Similarly, $ \text{RNN}_L$ also leverages the users' representations to update the representations of each POI $l$ at next timestamp $\mathbf{h}_l^{t+1}$. Specifically, we denote the list of users who visited $l$ at timestamp $t$ as $\mathcal{U}(l,t)$, and denote the list of their representations as $\mathbf{h}_{\mathcal{U}(l,t)}^t$. The representation of POI $l$ is updated as: $\mathbf{h}_l^{t+1} = \text{RNN}_L(\mathbf{h}_l^{t},\mathbf{h}_{\mathcal{U}(l,t)}^{t})$.
At each timestamp $t$, the representation of POI $l$ is recursively updated $|\mathcal{U}(l,t)|$ times in the chronological order of the users' visits. 
To update the user-POI representations in parallel while maintaining the time dependency, we create each mini-batch by selecting the independent user-POI visits from the users’ trajectories, to ensure that every two visits processed in the same batch do not share any common users or POIs \cite{kumar2019predicting}.

\begin{table}[t]
\caption{Summary statistics of datasets.}
\vspace{-4mm}

\label{tb:1}
\scalebox{0.82}{
\centering
\begin{tabular}{lccccccccc}
\hline
 &
  \#Users &
   &
  \#POIs &
   &
  \#Visits &
   &
  \begin{tabular}[c]{@{}l@{}}\#Family \\Relations\end{tabular} &
   &
  \begin{tabular}[c]{@{}l@{}}\#Colleague \\Relations\end{tabular} \\ \hline
B'more July & 17,892 &  & 559  &  & 1,753,382 &  & 19,026 &  & 17,646 \\
DC July        & 7,517  &  & 4,213 &  & 262,679  &  & 1,100  &  & 12,316 \\
DC August      & 3,418  &  & 3,450 &  & 277,093  &  & 1,290  &  & 14,250 \\ \hline
\vspace{-3mm}
\end{tabular}
}
\vspace{-4mm}
\end{table}

\begin{table}[]
\caption{Comparison of  next POI recommendation methods.}
\vspace{-4mm}

\scalebox{0.87}{
\begin{tabular}{l|l|l|l|l}
\hline
&     & User & Social &  \\
& Model   &  Embeddings & Relations & Trajectories \\ \hline
\begin{tabular}[c]{@{}l@{}}Statistical  \\ Method\end{tabular}                    & 
Markov \cite{mathew2012predicting}  &                 &           &   \hspace{2em} \checkmark          \\ \hline
\begin{tabular}[c]{@{}l@{}}Graph\\ Network\end{tabular}                           & 
LightGCN  \cite{he2020lightgcn} &   \hspace{2em} \checkmark             &\hspace{1.5em}\checkmark         &              \\ \hline
\multirow{2}{*}{\begin{tabular}[c]{@{}l@{}}Recurrent\\ Network\end{tabular}}      & 
GRU4Rec   \cite{hidasi2015session} &                 &           & \hspace{2em} \checkmark     \\
                                                                                  & 
                                                                                  
DeepMove \cite{feng2018deepmove} &                 &           & \hspace{2em} \checkmark      \\ \hline
\multirow{3}{*}{\begin{tabular}[c]{@{}l@{}}Self-\\attentive \\ Method\end{tabular}} & 
SASRec \cite{kang2018self}  &                 &           & \hspace{2em} \checkmark      \\
                                                                                  & 
TiSASRec \cite{li2020time} &                 &           & \hspace{2em} \checkmark    \\
                                                                                  & 
STAN \cite{luo2021stan}    & \hspace{2em}\checkmark           &           &          \hspace{2em} \checkmark    \\ \hline
\begin{tabular}[c]{@{}l@{}}\textbf{Our} \\ \textbf{Method}\end{tabular}                        & 
\textbf{MEMO}     &  \hspace{2em}\checkmark         &  \hspace{1.5em}\checkmark       & \hspace{2em} \checkmark     \\ \hline
\end{tabular}}
\label{tb:2}

\vspace{-4mm}
\end{table}

\begin{table*}[tbp]
\vspace{-4mm}
\caption{Performance of next POI recommendation for different methods.}
\vspace{-4mm}

\label{tb:3}
\scalebox{1.1}{
\centering
\begin{tabular}{lcccccccc}
\hline
 & \multicolumn{2}{c}{Baltimore July} &  & \multicolumn{2}{c}{DC July} &  & \multicolumn{2}{c}{DC August} \\ \cline{2-3} \cline{5-6} \cline{8-9} 
         & Recall@10                        & MRR@10 &  & Recall@10 & MRR@10 &  & Recall@10 & MRR@10 \\ \cline{1-6} \cline{8-9} 
Markov    &  $0.257\pm0.001$   &     $0.088\pm0.004$  &     & $0.471\pm0.012 $ & $0.224\pm0.012$         &        & $ 0.431\pm0.013$  &  $0.224\pm 0.012$          \\
LightGCN  &  $0.650 \pm0.044$   &     $0.392\pm0.011$  &     &  $0.609\pm0.021$ &    $0.291\pm0.034$     &        &  $ 0.622\pm0.019$ & $0.311\pm0.014$           \\

GRU4Rec   &  $0.657\pm0.009$  &     $0.384\pm0.006$  &   &   $0.617\pm0.019$   & $0.300\pm0.013$         &        & $0.624 \pm0.005$ & $0.306\pm0.005$          \\
DeepMove  &  $0.715 \pm0.031$   &     $0.402\pm0.003$  &     &  $0.660\pm0.007$ &    $0.333\pm0.006$     &        &  $ 0.648\pm0.014$ & $0.331\pm0.009$           \\

SASRec  &  $0.738\pm 0.029$    &     $0.415\pm0.003$ & &   $0.694 \pm0.019$  &  $0.335\pm0.006$       &        & $0.687 \pm0.006$ & $ 0.353 \pm 0.007$      \\
TiSASRec&  $0.836 \pm0.002$    &     $0.420\pm0.007$  & &  $0.760 \pm0.011 $  & $0.353\pm0.014$  &        & $0.763\pm0.006$ &  $0.379\pm0.003$          \\
STAN     &  $0.868\pm0.003$     &     $0.426 \pm0.004$  & &  $0.807 \pm 0.005$ &     $0.360\pm0.007$   &        & $0.816\pm 0.007$&   $0.426\pm0.004$         \\ \hline
MEMO  &  $\mathbf{0.891}\mathbf{ \pm0.004}$   &  $\mathbf{0.446 }\mathbf{ \pm 0.009}$& & $\mathbf{ 0.831}\mathbf{ \pm 0.002}$  &   $\mathbf{0.380}\mathbf{ \pm0.006}$       &        & $\mathbf{0.840}\mathbf{ \pm0.003}$   &   $\mathbf{0.435}\mathbf{ \pm0.008}$       \\ \hline
\end{tabular}
}
\vspace{-4mm}
\end{table*}

\vspace{-2mm}
\begin{figure*}[t]
\centering
    \begin{minipage}[t]{0.24\textwidth}
    
    \centering
    \includegraphics[width=3.8cm]{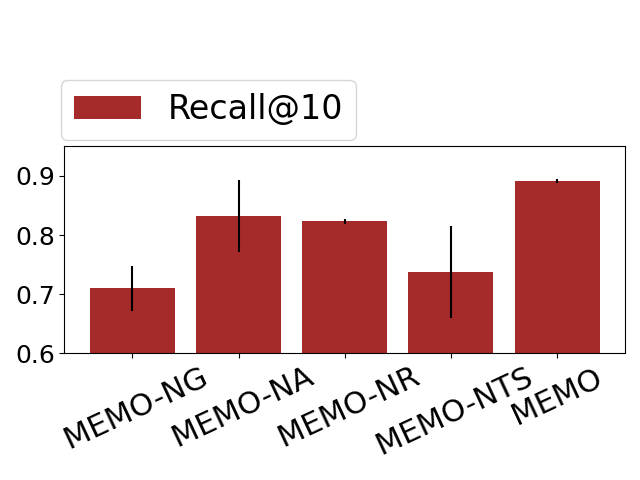}
    \vspace{-6mm}
    \caption*{Baltimore July}
    \label{}
    \end{minipage}
    \begin{minipage}[t]{0.24\textwidth}
    \centering
    \includegraphics[width=3.8cm]{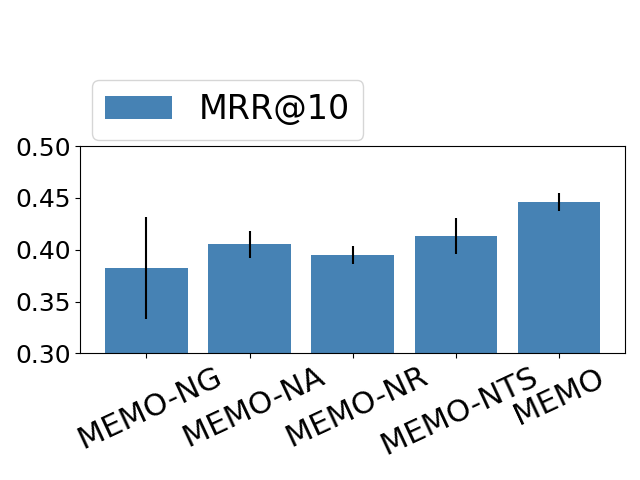}
    \vspace{-4mm}
    \caption*{Baltimore July}
    \label{}
    \end{minipage}
    \begin{minipage}[t]{0.24\textwidth}
    \centering
    \includegraphics[width=3.8cm]{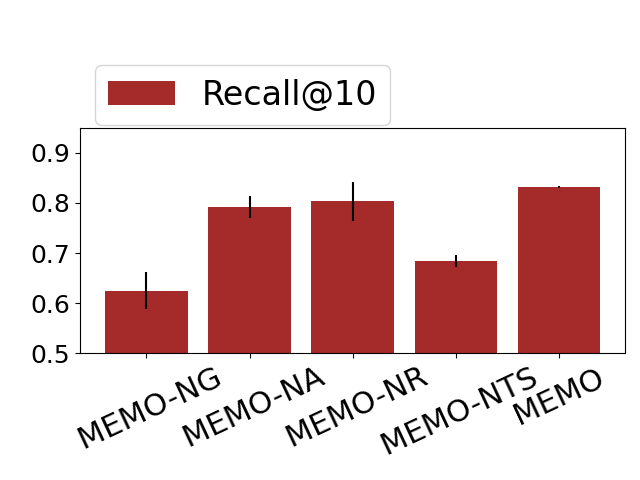}
    \vspace{-4mm}

    \caption*{DC July}
    \label{}
    \end{minipage}
    \begin{minipage}[t]{0.24\textwidth}
    \centering
    \includegraphics[width=3.8cm]{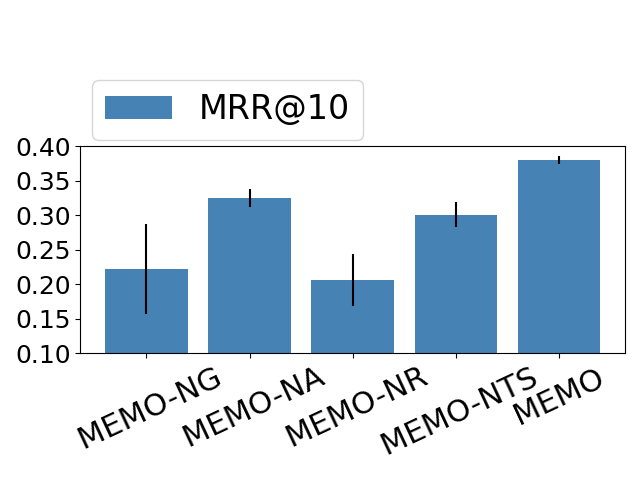}
    \vspace{-4mm}

    \caption*{DC July}
    \label{}
    \end{minipage}
  
    \vspace{-4mm}

  \caption{Ablation studies.}
  \label{fig:abl}
\end{figure*}

\begin{figure*}[t]
\centering
    \begin{minipage}[t]{0.24\textwidth}
    \centering
    \includegraphics[width=4cm, height =2.5cm]{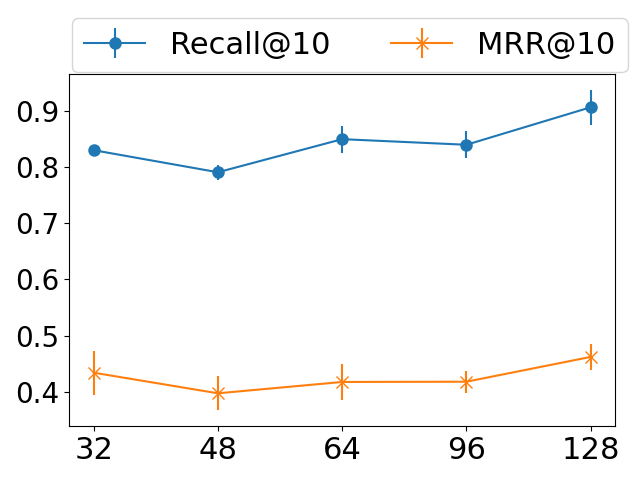}
    \vspace{-4mm}

    \caption*{Dimension}
    \end{minipage}
    \begin{minipage}[t]{0.24\textwidth}
    \centering
    \includegraphics[width=4cm, height = 2.5cm]{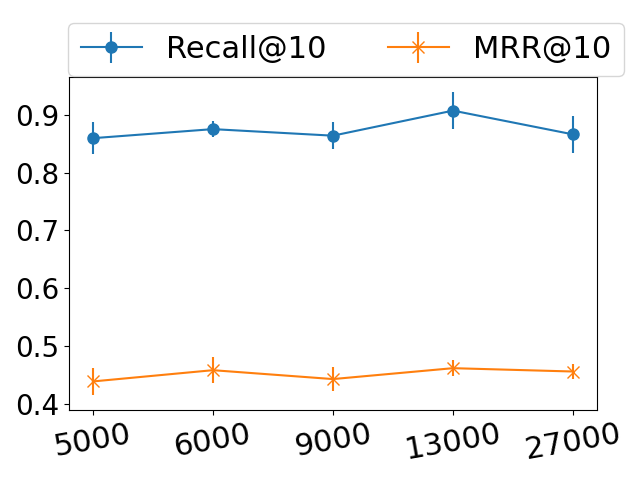}
    \vspace{-4mm}

    \caption*{\negthinspace Timestamp \negthinspace interval}
    \end{minipage}
        \begin{minipage}[t]{0.24\textwidth}
    \centering
    \includegraphics[width=4cm, height = 2.5cm]{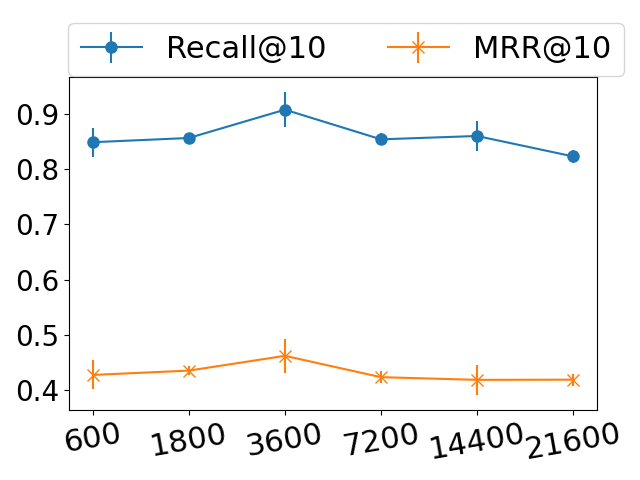}
        \vspace{-4mm}

    \caption*{Time threshold}
    \end{minipage}
        \begin{minipage}[t]{0.24\textwidth}
    \centering
    \includegraphics[width=4cm, height = 2.5cm]{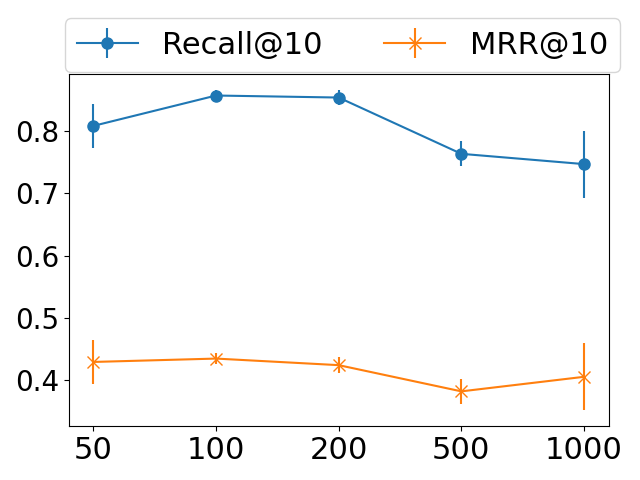}
    \vspace{-4mm}

    \caption*{Distance threshold}
    \end{minipage}
    
  \vspace{-4mm}
  \caption{Parameter studies.}
  \label{fig:param}
\end{figure*}

\noindent\textbf{Spatio-Temporal Representation.}
Apart from the users' relations and trajectories, we also incorporate the spatio-temporal information 
to facilitate next POI recommendation. For instance, when roaming around, a user might be interested in visiting a club near his/her current location (spatial) within the next few hours (temporal). We hence design a mechanism to efficiently incorporate such spatio-temporal information by utilizing the spatial transition (distance) and time transition (visiting time intervals) between each pair of POIs consecutively visited by a user. 
Specifically, we introduce two learnable vectors, $\mathbf{t}_s$ and $\mathbf{t}_l$ to  respectively represent the short and long time transitions. The representation of a time interval is then formally obtained as: 
$
    \mathbf{z}^{\triangle t}= 
\mathbf{t}*\mathbf{t} + tanh(\mathbf{t}),
$
where $\mathbf{t} = \mathbf{t}_s$ if the time transition is shorter than a preset threshold $\theta_t$, otherwise $\mathbf{t} = \mathbf{t}_l$.
Here, $*$ denotes the element-wise multiplication. 
Similarly, we obtain the representation for spatial intervals $\mathbf{z}^{\triangle d}$ 
with a spatial threshold $\theta_d$. The distance between two POIs is formed by the Haversine distance \cite{van2014heavenly}. After updating each user $u$'s representation $\mathbf{h}_u^t$ and each POI $l$'s representation $\mathbf{h}_l^t$, we recommend a list of POIs for user $u$ to visit at the next timestamp $t+1$ by selecting the POIs with the highest prediction scores computed from a fully connected layer. We model the prediction of the next POI as a multi-class classification problem, where the ground truth is the POI that each user actually visits at the next timestamp. 
We also adopt the cross-entropy loss for the model optimization \cite{feng2018deepmove}.

\section{Experiments}

We now evaluate the effectiveness of the proposed method in next POI recommendation by leveraging a real-world location data. 

\vspace{-3mm}
\subsection{Experimental Setting}

\noindent\textbf{Data.} 
 The population-scale individual-level location data are provided by a leading data aggregator, who aggregates the data from more than 400 commonly used mobile apps, via a proprietary software development kit (SDK) installed on these apps to minimize battery drainage while tracking locations. The data are collected in compliance with privacy regulations, including General Data Protection Regulation (GDPR) and California Consumer Privacy Act (CCPA). The data cover one quarter of the U.S. population and are representative of the population. Each user is tracked on average every five minutes.
 Each data record contains an anonymized user ID common across all apps, timestamp, longitude, and latitude of the location visited, speed, and dwell time. Specifically, we analyze three datasets over different time periods in 2019: Baltimore (B'more) July, DC July, and DC August. Analyzing the users' trajectories via the state-of-the-art stop point detection method, Infostop \cite{aslak2020infostop}, we detect the users' home and work locations as the most frequent stop points during the weekday nighttime and daytime respectively; and as a result infer their focal social relations, families and colleagues. 
Table \ref{tb:1} displays the summary statistics.



\noindent\textbf{Baselines.}
We compare the proposed framework with the representative baselines across four categories (Table \ref{tb:2}), including: 1) statistical methods, where the \textbf{Markov} \cite{mathew2012predicting} method recommends the next POIs by 
utilizing the dependency between every two consecutive visits; 2) Graph Network method, where \textbf{LightGCN} \cite{he2020lightgcn} leverages the user-POI relation graph and a simplified GCN model with controllable oversmoothing for recommendations;  3) Recurrent Network methods, where \textbf{GRU4Rec} \cite{hidasi2015session} and \textbf{DeepMove} \cite{feng2018deepmove} use RNNs to capture the users' preferences hidden in the trajectories; and 4) {Self-attentive methods}, including \textbf{SASRec} \cite{kang2018self}, \textbf{TiSASRec} \cite{li2020time}, and \textbf{STAN} \cite{luo2021stan}, which use self-attention to capture the long-term dependencies in the users' trajectories.

\noindent\textbf{Parameter Settings.}
For each dataset, we partition the training$/$ validation$/$ test set into $80\%/10\%/10\%$; and set the training epoch as 50 and learning rate as $0.001$. All reported results include the mean and standard deviation over 10 repeated runs. We use the Adam optimizer. By default, we set the representation dimension $d=128$, length of each of the 400 timestamps as $13,000$ seconds, time threshold $\theta_t=3600$ seconds, and distance threshold $\theta_d = 200$ meters. We use two popular metrics, Recall@K and MRR@K, to gauge the model performance, where K is set to 10.
\vspace{-2mm}
\subsection{Performance of Next POI Recommendation}
Table \ref{tb:3} compares the performance of the different methods. 
The primary findings include: 1) GRU4Rec and DeepMove consistently outperform the Markov chain model, as the latter focuses merely on the consecutive visits, whereas the RNN-based methods capture more patterns hidden in the trajectories; 2) LightGCN performs worse than the RNN-based models, suggesting that only utilizing the user-POI relations and static embedding limit the POI recommendation performance; 
3) STAN exhibits a competitive performance, as it learns the user embeddings for personalized recommendations, and explicitly leverages the spatio-temporal information with the self-attention mechanism; and 4) MEMO consistently outperforms all baselines, primarily because i) MEMO captures the users' preferences more comprehensively by integrating the users' representations learned from the different relations; ii) MEMO models the user-POI mutual influence over time, with the learned representations better capturing the latent status of the users/POIs.

\subsection{Ablation Studies}
We further conduct ablation studies to investigate the contributions of the different model components. Specifically, we compare MEMO with its variants: (a) MEMO-NG in which the relation modeling component is removed; 
(b) MEMO-NA in which the self-attention mechanism is replaced with an ordinary attention mechanism based on a bilinear function; 
(c) MEMO-NR in which the mutual influence component is removed, i.e., the representations are learned just from  the relation modeling component; and (d) MEMO-NTS in which the spatio-temporal representation is not used.
    
Given limited space, we only report the results on the Baltimore July and DC July data. The DC August data produce similar findings. Fig.~\ref{fig:abl} shows that MEMO-NG cannot achieve a satisfactory performance as it does not utilize the relational information. The performance of MEMO-NA also degrades as it implicitly assumes that all relation-specific representations are within the same latent space. The performance gap between MEMO and MEMO-NR indicates the importance of mutual influence modeling; and the comparison between MEMO-NTS and MEMO indicates the contribution of the spatio-temporal information to the recommendations.

\subsection{Parameter Studies}

To evaluate the robustness of our framework, we compare the performance under different hyperparameters. Fig.~\ref{fig:param} shows the performance of MEMO with the varied representation dimension $d$ in the range of $\{32, 48, 64, 96, 128\}$, 
the interval between two timestamps from $(5\sim 27)\times 10^3$ seconds, 
the time threshold $\theta_t$ from $600$ to $21600$ seconds, and distance threshold $\theta_d$ from $50$ to $1000$ meters. 
We will illustrate the results with the Baltimore July data. The findings from the other two data sets remain similar. In general, the model is not very sensitive to the parameter settings; yet the performance can still benefit from a proper fine-tuning.

\section{Conclusion}

In this research, we focus on next POI recommendation by addressing two main challenges: modeling different types of relations and user-POI mutual influence over time. Methodologically, we propose a novel framework, MEMO, to address these challenges by leveraging a multi-relational representation learning module to model and aggregate different relations, and further enlisting the coupled RNNs to capture the inter-temporal user-POI mutual influence. 
The extensive experiments conducted on the population-scale, yet fine-grained individual-level location big data, supported by comprehensive model comparisons, ablation and parameter studies, validate the superiority of the proposed framework. 

\bibliographystyle{ACM-Reference-Format}
\bibliography{ref}
\end{document}